\documentclass[twocolumn,showpacs,preprintnumbers,amssymb]{revtex4}
\usepackage{graphicx}
\usepackage{dcolumn}
\usepackage{bm}

\begin{document}
\preprint{CCOC-02-01}
\title{Periodic Phase Synchronization in coupled chaotic oscillators} 
\author{Won-Ho Kye$^1$} 
\email{whkyes@empal.com}
\author{Dae-Sic Lee$^{1,2}$}
\author{Sunghwan Rim$^1$}
\author{Chil-Min Kim$^1$}
\email{chmkim@mail.paichai.ac.kr}
\author{Young-Jai Park$^2$}
\affiliation{$^1$National Creative Research Initiative Center for Controlling Optical Chaos,\\
Pai-Chai University, Daejeon 302-735, Korea}
\affiliation{$^2$Department of Physics, Sogang University, Seoul 121-742, Korea}
\begin{abstract}
We investigate the characteristics of temporal phase locking states 
observed in the route to phase synchronization. 
It is found that before phase synchronization there is a periodic phase synchronization
state characterized by periodic appearance of temporal phase locking state and that
the state leads to local negativeness in one of the vanishing Lyapunov exponents.
By taking a statistical measure, we present the evidences of the phenomenon 
in unidirectionally and mutually coupled chaotic oscillators, respectively.
And it is qualitatively discussed that the phenomenon is described   
by a nonuniform oscillator model in the presence of noise.
\end{abstract}
%\narrowtext

\maketitle
%{\Large \bf Introduction}
		Over the past decade, synchronization in chaotic oscillators \cite{SyncOrg0,SyncOrg1} 
		has received much attention because of its fundamental importance 
		in nonlinear dynamics and potential
		applications to laser dynamics \cite{PSExp}, electronic circuits \cite{Circuit}, 
		chemical and biological systems \cite{NeuronSync},
		and secure communications \cite{Secure}.
		Synchronization in chaotic oscillators is characterized by the loss of exponential 
		instability in the transverse direction
		through interaction. 
		In coupled chaotic oscillators, it is known, various types of synchronization
		are possible to observe, among which are complete synchronization (CS) \cite{SyncOrg0,SyncOrg1}, 
		phase synchronization (PS) \cite{PhaseSync,Lee}, 
		lag synchronization (LS) \cite{Lag} 
		and generalized synchronization (GS) \cite{GenSync}.
			
		One of the noteworthy synchronization phenomena in this regard is PS 
		which is defined by the phase locking between nonidentical chaotic oscillators
		whose amplitudes remain chaotic and uncorrelated with each other: 
		%\begin{equation} 
		$|\theta_1-\theta_2| \leq  \mbox{const}.$ 
		%\end{equation}
		Since the first observation of PS in mutually coupled chaotic oscillators \cite{PhaseSync}, 
		there have been extensive studies in theory \cite{Lee}
		and experiments \cite{PSExp}.   
		The most interesting recent development in this regard is the report that
		the interdependence between physiological systems 
		is represented by PS and temporary phase-locking (TPL) states, e.g.,    
		(a) human heart beat and respiration \cite{HeartBeat}, 
		(b) a certain brain area and the tremor activity \cite{BrainMuscle}, 
		etc \cite{Visual, Wavelet}. 
		Application of the concept of PS in these areas sheds light 
		on the analysis of nonstationary bivariate data coming 
		from biological systems which was thought to be impossible 
		in the conventional statistical approach.
		And this calls new attention to the PS phenomenon. 
		
		Accordingly, it is quite important to elucidate a detailed 
		transition route to PS
		in consideration of the recent observation of a TPL state in 
		biological systems.
		What is known at present is that  TPL\cite{Lee} transits to PS
		and then transits to LS as the coupling strength increases. 
		On the other hand, it is noticeable that the phenomenon from   
		nonsynchronization to PS have hardly been studied, 
		in contrast to the wide observations of the TPL states in 
		the biological systems.

		The chief goal of this Letter is to study the characteristics of
                TPL states observed in the regime  
                from nonsynchronization to PS in coupled chaotic oscillators.
		We report that there exists a special locking regime in which
		a TPL state shows maximal periodicity, 
		which phenomenon we would call {\it periodic phase synchronization} (PPS).		
		We show this PPS state leads to local negativeness 
		in one of the vanishing Lyapunov exponents, taking the measure by which 
		we can identify 
		the maximal periodicity in a TPL state.
		We present a qualitative explanation of the phenomenon
		with a nonuniform oscillator model in the presence of noise.
                
		We consider here the unidirectionally coupled non-identical Ro\"ssler oscillators
                for first example:
                \begin{eqnarray}
                \dot{x}_{1}&=&-\omega_{1} y_{1} -z_{1},\nonumber\\
                \dot{y}_{1}&=& \omega_{1} x_{1} +0.15 y_{1},\nonumber\\
                \dot{z}_{1}&=& 0.2 +z_{1}(x_{1} -10.0), \nonumber\\ 
                \dot{x}_{2}&=&-\omega_{2} y_{2} -z_{2},\nonumber\\
                \dot{y}_{2}&=& \omega_{2} x_{2} +0.165 y_{2}+\epsilon(y_{1}-y_{2}),\nonumber\\
                \dot{z}_{2}&=& 0.2 +z_{2}(x_{2} -10.0), 
                \end{eqnarray}
                where the subscripts imply the oscillators 1 and 2, respectively, 
                $\omega_{1,2}$ $(=1.0 \pm 0.015)$ is the overall 
                frequency of each oscillator, and $\epsilon$ is the coupling strength.  
                It is known that PS appears in the regime $\epsilon \geq \epsilon_c$ 
                and that $2\pi$ phase jumps arise when $\epsilon < \epsilon_c$.
                Lyapunov exponents play an essential role in the investigation of
                the transition phenomenon with coupled chaotic oscillators 
                and as generally understood that PS transition
                is closely related to the transition to the negative value 
		in one of the vanishing Lyapunov exponents \cite{SyncOrg1}.

                Figure 1 shows two largest conditional Lyapunov
                exponents from Eq. (1) according to the coupling strength $\epsilon$.
                One can see the dip characterized by the local negativeness
                in the vanishing Lyapunov exponent.
                The reference points A and C indicate the borders of the dip and 
                B its center.
                PS transition occurs in the right of C ($\epsilon=0.085$) 
                where the phase difference
                of coupled chaotic oscillators is bounded within a constant. 
                The temporal behaviors at the three reference points are presented in Fig. 2.
                The phase jumps at B 
                looks quite regular, compared to those of A and C.
                Though this observation may seem rather intuitive, 
                we shall see that it is a valid one and that all the phenomenon 
                is deeply related to the dip, the local negativeness 
                in the vanishing Lyapunov exponent of Fig. 1.

\begin{figure}
\begin{center}
\rotatebox[origin=c]{0}{\includegraphics[width=8.5cm]{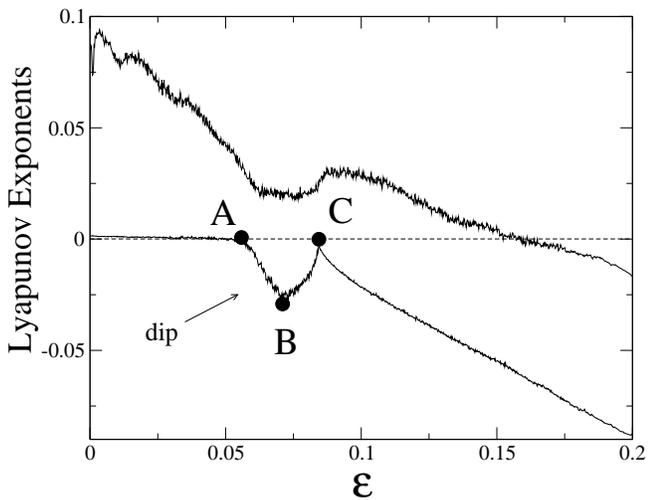}}
\caption{Two largest conditional Lyapunov exponents and 
	the dip in coupled R\"ossler oscillators when 
	$\omega_0=1.0$: 
	A ($\epsilon=0.058$), B($\epsilon=0.072$), and  C ($\epsilon=0.083$) are the reference 
	points to explain different characteristics of phase dynamics.}
\end{center}
\end{figure}
		A vanishing Lyapunov exponent corresponds to a phase variable of an oscillator
                and it exhibits the neutrality of an oscillator in the phase direction. 
                Accordingly, the local negativeness of an exponent 
                indicates this neutrality is locally broken \cite{PhaseSync}. 
		It is important to define an appropriate phase variable
		in order to study the TPL state more thoroughly.
		In this regard, several methods have been proposed  
		methods of using linear interpolation at a Poincar\'e  
		section \cite{PhaseSync}, 
		phase space projection \cite{PhaseSync,Lee},  
		tracing of the center of rotation in phase space \cite{Spline}, 
		Hilbert transformation \cite{PhaseSync,Wavelet}, or wavelet transformation \cite{Wavelet}. 
		Among these we take the method of phase space projection
		onto the $x_1-y_1$ and $x_2-y_2$ planes 
		with the geometrical relation $\theta_{1,2}=\arctan(y_{1,2}/x_{1,2})$, 
		and obtain phase difference $\phi=\theta_1-\theta_2$.

		The system of coupled oscillators is said to be in a TPL state
		(or laminar state) when  
		$\langle \dot{\phi} \rangle < \Lambda_c$ 
		where $\langle ... \rangle$ is the running average over appropriate short time scale and
		$\Lambda_c$ is the cutoff
		value to define a TPL state.
		The locking length of the TPL state, $\tau$,  is defined by time interval between
		two adjacent peaks of $\langle \dot{\phi} \rangle$ (see Fig. 2).
\begin{figure}
\begin{center}
\rotatebox[origin=c]{0}{\includegraphics[width=8.5cm]{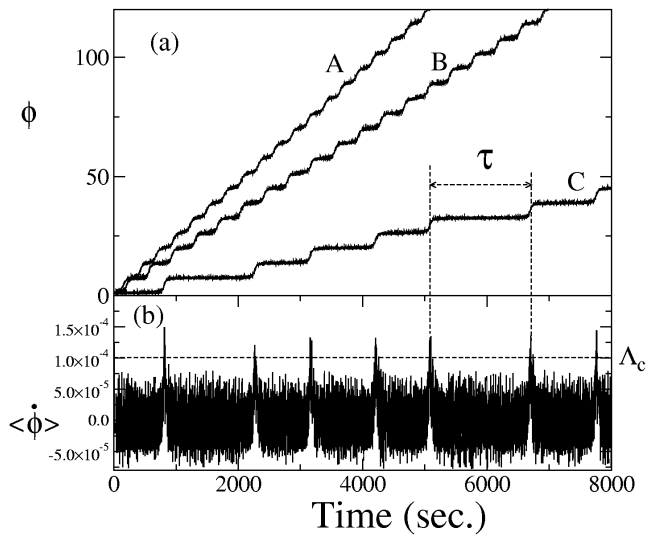}}
\caption{Temporal behaviors of the phase difference:
	(a) $\phi$ at each reference point. 
	$\tau$ is the locking length of a TPL state.
	(b) $\langle \dot{\phi} \rangle$ at reference point C 
	(running averaged over 10 seconds 
	to remove irrelevant fast fluctuations). 
	The cutoff value for TPL a state is $\Lambda_{c}=1.0\times 10^{-4}$.} 
\end{center}
\end{figure}
		In order to study the characteristics 
		of the locking length $\tau$, 
		we introduce a measure:
		\begin{eqnarray}
			P(\epsilon)= \frac{\sqrt{\mbox{var}(\tau)}}{\langle \tau \rangle},
		\end{eqnarray}
		which is the ratio between the average value of 
		time lengths of TPL states and their standard deviation.
		In terminology of stochastic resonance, 
		it can be interpreted as noise-to-signal ratio \cite{Coherence,SR}.
		The measure would be minimized 
		where the periodicity is maximized in TPL states.
\begin{figure}
\begin{center}
\rotatebox[origin=c]{0}{\includegraphics[width=8.5cm]{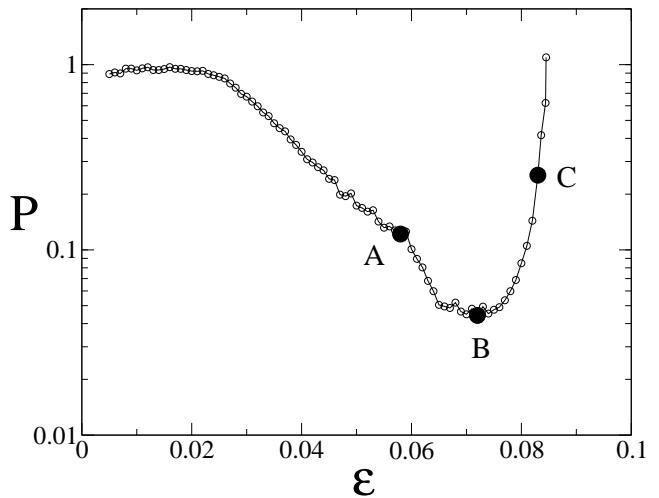}}
\caption{Measure P as a function of $\epsilon$. Big dots 
	are the reference points of Fig.1 and 
	each point is the average value of 2500 TPL states.} 
\end{center}
\end{figure}

	Measure $P$ as a function of $\epsilon$ is presented in Fig. 3.
	The big dots indicate the reference points of Fig. 1.
	We see that the value of P begins to drop rapidly from reference
	point A which corresponds to the left border of the dip in Fig. 1. 
	And the value of $P$ is minimized in a broad region around B, the center of the dip.
	The value rapidly increases 
	after passing reference point C, the right border of the dip. 
	What is interesting here is that
	in the region from $\epsilon=0.062$ to $0.078$ the periodicity
	is maximized and corresponds to the central part of the dip. 	
	Eventually, the coupled chaotic oscillators develop to PS near $\epsilon=0.085$. 
	The result presented in Fig. 3 leads us to argue that 
	the dip in a vanishing Lyapunov exponent
	shows the maximal periodicity of TPL states around the minimum of $P$.
	We call the TPL state inside the dip a PPS state in
	the sense that the TPL state appears rather periodically 
	than outside the dip.
	 
	To validate the argument, we explain the phenomenon in simplified
	dynamics.  From Eq. (1), we obtain the equation of motion in terms 
	of phase difference:
\begin{eqnarray}
        \frac{d \phi}{dt}&=&\Delta \omega +A(\theta_1, \theta_2, \epsilon) 
			\sin\phi + \xi(\theta_1, \theta_2, \epsilon),
\end{eqnarray}
        where,
\begin{eqnarray}
        A(\theta_1, \theta_2, \epsilon) &=& (\epsilon+0.15)\cos(\theta_1+\theta_2)
				-\frac{\epsilon}{2}(\frac{R_1}{R_2}), \nonumber\\
        \xi(\theta_1, \theta_2, \epsilon) &=& 
	\frac{\epsilon}{2}\frac{R_1}{R_2}\sin(\theta_1+\theta_2)
	 + \frac{z_1}{R_1}\sin(\theta_1)-\frac{z_2}{R_2}\sin(\theta_2) \nonumber \\
	 & & +(\epsilon + 0.015) \cos(\theta_2) \sin(\theta_2).\nonumber
\end{eqnarray}
Here $\Delta \omega=\omega_1-\omega_2$ and $R_{1,2}= \sqrt{x_{1,2}^2 + y_{1,2}^2}$. 
And from Eq. (3) we obtain the simplified equation to describe the phase dynamics: 
${d \phi }/{dt} = \Delta \omega +\langle A \rangle \sin(\phi) + \xi$,
where $\langle A \rangle$ is the time average of $A(\theta_1, \theta_2, \epsilon)$. 
This is a nonuniform oscillator in the presence of noise 
where $\xi$ plays a role of effective noise \cite{Strogatz} and 
the value of $\langle A \rangle$ controls the width of bottleneck (i.e, nonuniformity of the flow).
If the bottleneck is wide enough, 
(i.e., faraway from the saddle-node bifurcation point: $\Delta \omega \gg -\langle A \rangle$),
the effective noise hardly contributes to the phase dynamics of the system. 
So the passage time is wholly governed by the width of the bottleneck as follows:
$\langle \tau \rangle \sim 1/\sqrt{\Delta \omega ^2 -\langle A \rangle ^2} 
	\sim 1/\sqrt{\Delta\omega^2 - \epsilon^2/4}$,
which is a slowly increasing function of $\epsilon$.
In this region while the standard deviation of TPL states is nearly constant 
(because the widely opened bottlenecks periodically appears and those lead to
small standard deviation), the average value of locking length of TPL states is relatively short
and the ratio between them is still large.
Accordingly, the value of $P(\epsilon)$ slowly decreases 
in the regime before reference point B in Fig. 3.

On the contrary  as the bottleneck becomes narrower 
(i.e., near the saddle-node bifurcation point: $\Delta \omega \geq -\langle A \rangle$)
the effective noise begins to perturb the process of bottleneck passage
and regular TPL states develop into intermittent ones (see C in Fig. 2 (a)) \cite{Lee,Kye}. 
It makes the standard deviation increase very rapidly and 
this trend overpowers that of the average value 
of locking lengths of the TPL states. 
For that reason, the value of $P(\epsilon)$ rapidly increases 
passing the PPS regime in Fig. 3.
Thus we understand that the competition between width of bottleneck
and amplitude of effective noise produces the crossover 
at the minimum point of $P(\epsilon)$ 
which shows the maximal periodicity of TPL states.

\begin{figure}
\begin{center}
\rotatebox[origin=c]{0}{\includegraphics[width=8.5cm]{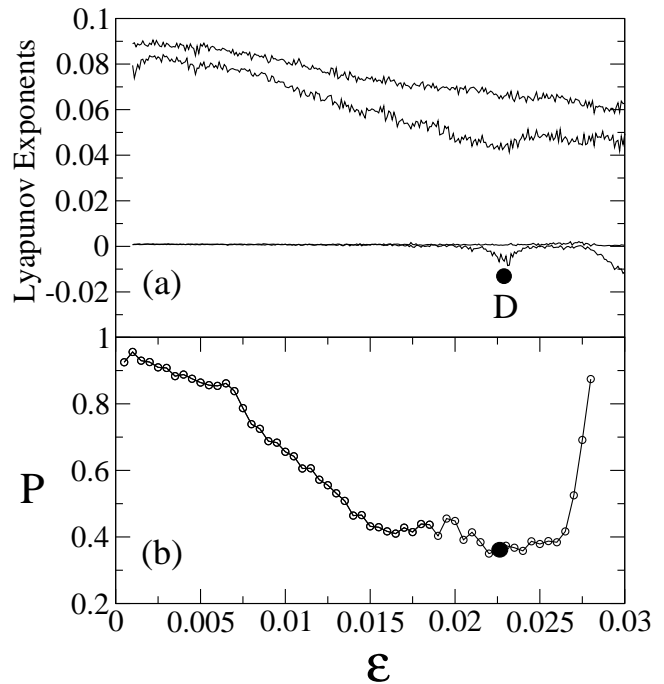}}
\caption{ Lyapunov exponents and measure P in mutually coupled R\"ossler oscillator:
	$\dot{x}_{1,2}=-\omega_{1,2} y_{1,2} -z_{1,2}+\epsilon(x_{2,1}-x_{1,2}),
       	\dot{y}_{1,2}= \omega_{1,2} x_{1,2} +0.15 y_{1,2},
       	\dot{z}_{1,2}= 0.2 +z_{1,2}(x_{1,2} -10.0)$, where $\omega_{1,2}=1.0\pm0.015$.
	(a) One can see the dip in one of the vanishing Lyapunov exponents
	and the reference point D 
	indicates the center of it.
	(b) The big dot shows the minimum point and it coincides with the point of 
	the center of the dip.  Each point is the average value of 12000 TPL states.}
\end{center}
\end{figure}

Rosenblum et al. firstly observed the dip in mutually coupled chaotic oscillators \cite{PhaseSync}. 
However the origin and the dynamical characteristics of the dip
have been left unclarified. We argue that the dip observed in mutually coupled
chaotic oscillators has the same origin as observed above in unidirectionally coupled systems.
Figure 4 shows the four largest Lyapunov exponents and the value of $P$ according 
to the coupling strength $\epsilon$.
We can see that the minimum region of $P$ from $\epsilon=0.022$ to $0.027$ 
coincides with the central part of the dip (D in the figure)
even though the valley in Fig. 4 (b) is not deeper than that of the unidirectionally coupled systems.
Thus it is reasonably summed up that a PPS state exists just before the transition 
to PS and that the dip observed by Rosenblum et al. is the very evidence of a PPS state
in mutually coupled chaotic oscillators.

Common apprehension is that
near the border of synchronization the phase difference in coupled regular 
oscillators is periodic \cite{PhaseSync} 
whereas in coupled chaotic oscillators 
it is irregular \cite{Lee}.
On the contrary, we report that the special
locking regime exhibiting the maximal periodicity of a TPL
state also exists in the case of coupled chaotic oscillators. 
In general, the phase difference of coupled chaotic oscillators is
described by the one-dimensional Langevin equation:
$\dot{\phi}=F(\phi) + \xi$ where $\xi$ is the effective noise with finite amplitude.
The investigation with regard to PS transition is the study of scaling of the laminar
length around the virtual fixed point $\phi^*$ where $F(\phi^*)=0$ \cite{Kye,Virtual} and
PS transition is established when $|\int^{\phi^*}_{\phi} F(\phi) d\phi| > \max|\xi|$. 
Consequently, the crossover region, from which the value of $P$ grows exponentially 
(as shown in Fig. 3-4), exists 
because intermittent series of TPL states with longer locking length $\tau$ 
appears as PS transition is nearer.
Eventually it leads to an exponential growth of 
the standard deviation of the locking length.
Thus we argue that PPS is the generic phenomenon 
mostly observed in coupled chaotic 
oscillators prior to PS transition.

In conclusion, analyzing the dynamic behaviors in coupled chaotic oscillators with
slight parameter mismatch we have completed the whole transition route to PS. 
We find that there exists a special locking regime called PPS in which
a TPL state shows maximal periodicity and
that the periodicity leads to local negativeness in 
one of the vanishing Lyapunov exponents.
We have also made a qualitative description of this phenomenon
with the nonuniform oscillator model in the presence of noise.
Investigating the characteristics of TPL states between nonsynchronization and PS, 
we have clarified the transition route before PS.
Since PPS appears in the intermediate regime 
between non-synchronization and PS,
we expect that the concept of PPS can be used as a tool for analyzing 
weak interdependences, i.e. those not strong enough to develop to PS,
between nonstationary bivariate data coming from biological systems, for instance.
Moreover PPS could be a possible mechanism of the 
chaos regularization phenomenon \cite{Regular,Regular1} 
observed in neurobiological experiments. 

{\it Note added.} - Recently, we were informed by S. Boccaletti that 
		    the phenomenon observed by us was confirmed 
		    in CO$_2$ laser systems, experimentally \cite{Bocca}.

The authors acknowledge the correspondence with S. Boccaletti, E. Allaria, R. Meucci, and F.T. Arecchi
about experimental confirmation of PPS and thank 
		A. Pikovsky, Y. -C. Lai, K. Josi\'c, and M. Choi for helpful comments.
		This work is supported by Creative Research Initiatives of 
		the Korean Ministry of Science and Technology.

\end{document}